\documentclass[preprint]{aastex63}

\usepackage[utf8]{inputenc}
\usepackage{amsmath, color, amssymb, latexsym, graphicx, lipsum}

\newcommand{\ua}{\textbf{u}_{\text{A}}}
\newcommand{\da}{\boldsymbol\nabla\cdot\textbf{u}_{\text{A}}}
\newcommand{\dap}{\boldsymbol\nabla'\cdot\textbf{u}_{\text{A}}'}
\newcommand{\uh}{\textbf{u}}
\newcommand{\dv}{\boldsymbol\nabla\cdot\textbf{u}}
\newcommand{\dvp}{\boldsymbol\nabla'\cdot\textbf{u}'}
\submitjournal{ApJ}
\shorttitle{Compressible solar wind turbulence}
\shortauthors{Andr\'es et al.}

\begin{document}

\title[]{The evolution of compressible solar wind turbulence in the inner heliosphere: PSP, THEMIS and MAVEN observations}

\correspondingauthor{Nahuel Andr\'es}
\email{nandres@iafe.uba.ar, nandres@df.uba.ar}

\author[0000-0002-1272-2778]{N. Andr\'es}
\affiliation{Instituto de Astronom\'ia y F\'{\i}sica del Espacio, CONICET-UBA, Ciudad Universitaria, 1428, Buenos Aires, Argentina}
\affiliation{Departamento de F\'{\i}sica, Facultad de Ciencias Exactas y Naturales, UBA, Ciudad Universitaria, 1428, Buenos Aires, Argentina}

\author{F. Sahraoui}
\affiliation{Laboratoire de Physique des Plasmas, \'Ecole Polytechnique, CNRS, Sorbonne University, Observatoire de Paris, Univ. Paris-Sud, F-91128 Palaiseau Cedex, France}

\author[0000-0002-8587-0202]{L. Z. Hadid}
\affiliation{Laboratoire de Physique des Plasmas, \'Ecole Polytechnique, CNRS, Sorbonne University, Observatoire de Paris, Univ. Paris-Sud, F-91128 Palaiseau Cedex, France}

\author{S.Y. Huang}
\affiliation{School of Electronic and Information, Wuhan University, Wuhan, China}

\author[0000-0001-9210-0284]{N. Romanelli}
\affiliation{Solar System Exploration Division, NASA Goddard Space Flight Center, Greenbelt, MD, USA}
\affiliation{CRESST II, University of Maryland, Baltimore County, Baltimore, MD, USA.}

\author[0000-0001-8685-9497]{S. Galtier}
\affiliation{Laboratoire de Physique des Plasmas, \'Ecole Polytechnique, CNRS, Sorbonne University, Observatoire de Paris, Univ. Paris-Saclay, F-91128 Palaiseau Cedex, France}
\affiliation{Institut universitaire de France, France}

\author[0000-0002-2778-4998]{G. DiBraccio}
\affiliation{Solar System Exploration Division, NASA Goddard Space Flight Center, Greenbelt, MD, USA}

\author[0000-0001-5258-6128]{J. Halekas}
\affiliation{Department of Physics and Astronomy, University of Iowa, Iowa City, IA, USA}

\begin{abstract}
The first computation of the compressible energy transfer rate from $\sim$ 0.2 AU up to $\sim$ 1.7 AU is obtained using PSP, THEMIS and MAVEN observations. The compressible energy cascade rate $\varepsilon_C$ is computed for hundred of events at different heliocentric distances, for time intervals when the spacecraft were in the pristine solar wind. The observational results show moderate increases of $\varepsilon_C$ with respect to the incompressible cascade rate $\varepsilon_I$. Depending on the level of compressibility in the plasma, which reach up to 25 $\%$ in the PSP perihelion, the different terms in the compressible exact relation are shown to have different impact in the total cascade rate $\varepsilon_C$. Finally, the observational results are connected with the local ion temperature and the solar wind heating problem.
\end{abstract}

\section{Introduction}\label{sec:intro}

Turbulence is a unique nonlinear phenomenon in fluid and plasma flows that allows the transfer of the energy between different scales. Turbulence plays a major role in controlling the dynamical features of many astrophysical plasmas such as accretion disks, star formation, solar wind heating, or energy transport in planetary magnetospheres \citep[e.g.,][]{B1998,S2009,He2012,F2012,Ch2015,K2017,H2020a,H2020b,S1999}. Thanks to the availability of in situ measurements from various orbiting spacecraft, the solar wind provides a unique opportunity to investigate plasma turbulence \citep{S2020,Ch2020,H2020a,A2019b,H2019,H2017b,K2015,A2013,H2012,O2011}. A long-established challenge in the solar wind community is the so-called heating problem. It is manifested by the fact that the solar wind proton temperature decreases slowly as a function of the radial distance from the Sun, in comparison to the prediction of the adiabatic expansion model of the solar wind \citep{M1982c,V2007,Pi2020}. While several scenarios have been proposed to explain those observations \citep[see,][]{M1999,M1991} the main candidate is certainly the local heating of the solar wind plasma via the turbulent cascade \citep{B2005,M2011}. In this picture, the energy that is injected at the largest scales in the solar wind will cascade within the inertial range, until it reaches the dissipation scales, where it is eventually converted into the thermal heat of the plasma particles \citep[see,][]{S2020,K2015}. This framework has led to several investigations to estimate the energy cascade rate in the solar wind at different scales and different heliocentric distances using theoretical, numerical and observational strategies.

The first theoretical exact relation (or law) for incompressible hydrodynamics (HD) turbulence was derived from the von-K\'arm\'an-Howarth dynamical equation \citep{vkh1938} and represents one of the very few exact results in turbulence theories \citep{F1995}. Under the assumption of homogeneity and full isotropy, the so-called 4/5 law \citep{K1941a,K1941b} predicts a linear scaling for the longitudinal third-order structure function of the velocity field with the distance between points. This exact relation (valid only in the inertial range) gives an expression for the energy dissipation or cascade rate $\varepsilon$ as a function of the structure functions of the turbulent fields \citep[see, e.g. ][and references therein]{MY1975}. \citet{Ga2011} and \citet{B2014} generalized this exact result to compressible HD turbulence within the isothermal and polytropic approximations, respectively. The authors have found the presence of a new term that acts in the inertial range as a source (or a sink) for the mean energy cascade rate, in contrast with incompressible HD turbulence, where only flux terms act to transfer energy in the inertial range. Numerical results of supersonic isothermal HD turbulence have shown that these new source terms are smaller than the flux terms in the inertial range \citep{K2013}.

The first generalization of these exact relations to a magnetized plasma was made by \citet{P1998a,P1998b} using the incompressible magnetohydrodynamics (MHD) model. The validity of this exact result has been subjected to several numerical tests using direct numerical simulations (DNSs) of MHD turbulence \citep[see, e.g.][]{Mi2009,Bo2009,W2010}. Moreover, the exact law has been used to estimate the energy cascade rate \citep{SV2007,Sa2008,Co2015} and the magnetic and kinetic Reynolds numbers \citep{WEY2007} in solar wind turbulence, and in large scale modeling of the solar wind \citep{M1999,Mc2008}. \citet{B2013} derived an exact law for a two-point correlation functions of the fields for isothermal compressible MHD turbulence, and was expressed in terms of flux or source terms. Recently, \citet{A2017b} revisited that work and provided a new derivation using the classical plasma variables, i.e., the plasma density and velocity field and the compressible Alfvén speed. The new expression reported in \citet{A2017b} showed four types of terms that are involved in the nonlinear cascade: the hybrid and $\beta$-dependent terms, in addition to well-known flux and source terms \citep[see,][]{A2019,F2020}. It is this latter formulation that we shall use in the present study.

From the observational viewpoint, using a reduced form of the exact relation for compressible MHD turbulence and in situ measurements from the Time History of Events and Macroscale Interactions during Substorms (THEMIS) spacecraft \citep{Au2009}, \citet{B2016c} and \citet{H2017a} have studied the role of compressibility in the energy cascade of the solar wind turbulence. The authors found a more prominent role of density fluctuations in amplifying the energy cascade rate in the slow than in the fast solar wind. Another interesting feature that has been evidenced in the Earth's magnetosheath is that density fluctuations reinforce the anisotropy of the energy cascade rate with respect to the local magnetic field \citep{H2017b} and increase the cascade rate as it enters into the sub-ion scales \citep{A2019b}. Recently, using Parker Solar Probe (PSP) observations during the first encounter, \citet{B2020} have computed the incompressible energy transfer rate between 55 and 35 solar radius. Their findings showed that the incompressible energy cascade rate obtained near the first perihelion is about 100 times higher than the average value at 1 AU \citep[e.g.,][]{SV2007,H2017a}. Moreover, \citet{A2020} have computed the first estimation of the incompressible energy cascade rate at the MHD scales in the plasma upstream of the Martian bow shock (at $\sim 1.38-1.67$ AU). Using Mars Atmosphere and Volatile EvolutioN (MAVEN) observations, the authors found that the nonlinear cascade of energy at the MHD scales is slightly amplified when proton cyclotron waves are present in the plasma. However, in all those recent cases only the incompressible cascade rate was estimated.

The main goal of the present paper is to generalize and extend previous observational studies using a more complete theory of turbulence to investigate the compressible energy cascade rate $\varepsilon_\mathrm{C}$ at the MHD scales at different heliocentric distances. In particular, we use 3 data sets, the observations from PSP at $\sim$ 0.2 - 0.4 AU, THEMIS around $\sim$ 1 AU, and MAVEN at $\sim$ 1.5 - 1.7 AU. The paper is organized as follows: in section \ref{sec:theo} we describe the compressible MHD set of equations and recall briefly the main steps to derive the exact laws for fully developed compressible MHD turbulence (and its incompressible limit). In section \ref{sec:obs}, we present the three observations sets and the selection criteria used in the present work. Finally, in section \ref{sec:res} we discuss our main observational results and their physical implications on solar wind turbulence.

\section{Theoretical model}\label{sec:theo}

\subsection{Compressible MHD equations}\label{subsec:model}

The three-dimensional (3D) compressible MHD equations are the continuity equation for the mass density $\rho$, the momentum equation for the velocity field {\bf u}, in which the Lorentz force is included, the induction equation for the magnetic field {\bf B}, and the differential Gauss' law.  These equations can be written as \citep[see, e.g.,][]{M1987,A2017a},
\begin{align}\label{1} 
    & \frac{\partial e}{\partial t}  = - \uh\cdot\boldsymbol\nabla e-c_s^2\boldsymbol\nabla\cdot\uh , \\ \label{2}
	&\frac{\partial \textbf{u}}{\partial t} = -\uh\cdot\boldsymbol\nabla\uh  + \ua\cdot\boldsymbol\nabla\ua - \frac{1}{\rho}\boldsymbol\nabla(P+P_M) - \ua(\da) + \textbf{f}_k  + \textbf{d}_k , \\ 	\label{3} 
    &\frac{\partial \ua}{\partial t} = - \uh\cdot\boldsymbol\nabla\ua + \ua\cdot\boldsymbol\nabla\uh -\frac{\ua}{2}(\dv) + \textbf{f}_m + \textbf{d}_m , \\  \label{4} 
    &\ua\cdot\boldsymbol\nabla\rho + 2\rho(\da) = 0,
\end{align}
where we have defined the compressible Alfv\'en velocity $\ua\equiv\textbf{B}/\sqrt{4\pi\rho}$. In this manner, both field variables, $\uh$ and $\ua$, are expressed in speed units. For the sake of simplicity we assume that the plasma follows an isothermal equation of state, i.e., $P=c_s^2\rho$, where $c_s$ is the sound speed (temperature dependent), which allows us to close the hierarchy of the fluid equations (no energy equation is further needed). Note that $P_M\equiv\rho u_\text{A}^2/2$ is the magnetic pressure and that the continuity equation \eqref{1} is written as a function of the internal compressible energy for an isothermal plasma, i.e., $e \equiv c_s^2\ln(\rho/\rho_0)$, where $\rho_0$ is a constant (of reference) mass density. Finally, \textbf{f}$_{k,m}$ are respectively a mechanical and the curl of the electromotive large-scale forcings, and $\textbf{d}_{k,m}$ are respectively the small-scale kinetic and magnetic dissipation terms.

\subsection{Exact relation in compressible MHD turbulence}\label{subsec:law}

Using Eq.~\eqref{1}-\eqref{4} and following the usual assumptions for fully developed homogeneous turbulence (i.e., infinite kinetic and magnetic Reynolds numbers and a steady state with a balance between forcing and dissipation \citep{Ga2011,B2018}, an exact relation for compressible MHD turbulence can be obtained as,
\begin{equation}\label{exactlaw}
	-2\varepsilon_C=\frac{1}{2}\boldsymbol\nabla_\ell\cdot\textbf{F}_\text{C}+\text{S}_\text{C}+\text{S}_\text{H}+\text{M}_\beta,
\end{equation}
where $\varepsilon_C$ is the total compressible energy cascade rate and $\textbf{F}_\text{C}$, $\text{S}_\text{C}$, $\text{S}_\text{H}$ and M$_\beta$ represent the total compressible flux, source, hybrid and $\beta$-dependent terms, respectively \citep[for a detailed derivation see,][]{A2017b}. These terms are defined as,
\begin{align}\nonumber
	\textbf{F}_\text{C} \equiv&~ \langle [(\delta(\rho\uh)\cdot\delta\uh+\delta(\rho\ua)\cdot\delta\ua]\delta\uh - [\delta(\rho\uh)\cdot\delta\ua+\delta\uh\cdot\delta(\rho\ua)]\delta\ua\rangle \\ \label{term_flux}
	&+ 2\langle\delta e\delta\rho\delta\uh\rangle, \\ \nonumber
  \text{S}_\text{C} \equiv&~ \langle[R_E'-\frac{1}{2}(R_B'+R_B)](\dv)+[R_E-\frac{1}{2}(R_B+R_B')](\dvp)\rangle \\ \nonumber
	&+\langle[(R_H-R_H')-\bar{\rho}(\uh'\cdot\ua)](\da) \\ \label{term_source}
	&+[(R_H'-R_H)-\bar{\rho}(\uh\cdot\ua')](\dap)\rangle, \\ \nonumber
  \text{S}_\text{H} \equiv&~ \langle\big(\frac{P_M'-P'}{2}-E'\big)(\dv)+\big(\frac{P_M-P}{2}-E\big)(\dvp)\rangle  \\ \nonumber &+\langle H'(\da)+H(\dap)\rangle \\ \label{term_hybrid}  &+\frac{1}{2}\langle\big(e'+\frac{u_\text{A}}{2}^{'2}\big)\big[\boldsymbol\nabla\cdot(\rho\uh)\big]+\big(e+\frac{u_\text{A}}{2}^2\big)\big[\boldsymbol\nabla'\cdot(\rho'\uh')\big]\rangle, \\ \label{term_beta}
	\text{M}_\beta \equiv&~ -\frac{1}{2}\langle\beta^{-1'}\boldsymbol\nabla'\cdot(e'\rho\uh) + \beta^{-1}\boldsymbol\nabla\cdot(e\rho'\uh') \rangle,
\end{align}
where we have defined the total energy (i.e., the free energy) and the density-weighted cross-helicity per unit volume respectively as,
\begin{align}\label{energy}
	E(\textbf{x}) \equiv &~\frac{\rho}{2}(\uh\cdot\uh+\ua\cdot\ua) + \rho e , \\
	H(\textbf{x}) \equiv &~\rho(\uh\cdot\ua),
\end{align}
and their associated two-point correlation functions as,
\begin{align}
	R_E(\textbf{x},\textbf{x}') \equiv&~ \frac{\rho}{2}(\uh\cdot\uh'+\ua\cdot\ua') + \rho e'  ,\\
	R_H(\textbf{x},\textbf{x}') \equiv&~ \frac{\rho}{2}(\uh\cdot\ua'+\ua\cdot\uh'), \\
   R_B(\textbf{x},\textbf{x}')\equiv&~\frac{\rho}{2}(\ua\cdot\ua').
\end{align}
In all cases the prime denotes field evaluation at $\textbf{x}'=\textbf{x}+\boldsymbol\ell$ ($\boldsymbol\ell$ being the displacement vector) and the angular bracket $\langle\cdot\rangle$ denotes an ensemble average. It is worth mentioning that the properties of spatial homogeneity implies (assuming ergodicity) that the results of averaging over a large number of realizations can be obtained equally well by averaging over a large region of space for one realization \citep{Ba1953}. In analysis of spacecraft data this generally amounts to time averaging based on the Taylor hypothesis \citep[e.g.,][]{T1938a,T1938b,H2019,T2019}. We have introduced the usual increments and local mean definitions, i.e., $\delta\alpha\equiv\alpha'-\alpha$ and $\bar{\alpha}\equiv(\alpha'+\alpha)/2$ (with $\alpha$ any scalar {or vector} function), respectively. Finally, we recall that the derivation of the exact law \eqref{exactlaw} does not require the assumption of isotropy and that it is independent of the dissipation mechanisms {acting} in the plasma (assuming that the dissipation acts only at the smallest scales in the system) \citep[see also,][]{Ga2011,A2016b,A2016c}.

The quantity in Eq.~\eqref{term_flux} is associated with the energy flux, and is the usual term present in the exact law of incompressible turbulence \citep{P1998a}. This term is written as a global divergence of products of increments of different variables. It is worth mentioning that the total compressible flux \eqref{term_flux} is a combination of two terms of different nature, a Yaglom-like term,
\begin{align}\label{f1}
	\textbf{F}_\text{1C} \equiv&~ \langle [(\delta(\rho\uh)\cdot\delta\uh+\delta(\rho\ua)\cdot\delta\ua]\delta\uh - [\delta(\rho\uh)\cdot\delta\ua+\delta\uh\cdot\delta(\rho\ua)]\delta\ua\rangle,
\end{align}
which is the compressible generalization of the incompressible term \citep[see, ][]{P1998a,P1998b}, and a new purely compressible flux term, 
\begin{align}\label{f2}
	\textbf{F}_\text{2C} \equiv&~ 2\langle \delta\rho\delta e \delta\uh\rangle,
\end{align}
which is a new contribution to the energy cascade rate due to the presence of density fluctuations in the plasma \citep[see,][]{B2013,A2017b}.

The purely compressible source terms in Eq. \eqref{term_source}, i.e. those proportional to the divergence of the Alfv\'en and kinetic velocity fields (and involve the two-point correlation functions $R_E$, $R_B$ and $R_H$), may act as a source (or a sink) for the mean energy cascade rate in the inertial range. The hybrid term offers the freedom to be written either as a flux- or as a source-like term. However, when written as a flux-like term it cannot be expressed as the product of increments, as the classical flux in incompressible HD and MHD turbulence \citep{vkh1938,K1941a,K1941b,Ch1951,P1998a,P1998b} or their counterparts in Eq.~{(\ref{term_flux})}. The mixed $\beta$-dependent term (transformed into a flux-like term in \citet{B2013} under certain conditions) has no counterpart in compressible HD turbulence \citep{Ga2011,B2014} and cannot, in general, be expressed as purely flux or source term. It is worth recalling that these three type of terms can be estimated only using multi-spacecraft data and techniques, since them include local vector divergences \citep{A2019b}. However, numerical results for supersonic and subsonic for HD and MHD turbulence show that these terms are negligible in the inertial range \citep{K2013,A2018}. Therefore, in the present paper, to estimate the compressible energy cascade rate \eqref{exactlaw}, we shall consider only the flux terms \eqref{f1} and \eqref{f2}. 

Assuming statistical isotropy, we can integrate Eq.~\eqref{exactlaw} over a sphere of radius $\ell$ to obtain a scalar relation for isotropic turbulence. In compact form Eq.~\eqref{exactlaw} can be cast as,
\begin{equation}\label{numlaw}
	-\frac{4}{3} \varepsilon_{C} \ell =\text{F}_\text{1C}+\text{F}_\text{2C},
\end{equation}
where $\text{F}_\text{C1}+ \text{F}_\text{C2} \equiv (\textbf{F}_\text{C1}+\textbf{F}_\text{C2})\cdot\boldsymbol{\bf\hat V}_{sw}$ is the flux term projected into the mean plasma flow velocity field ${\bf V}_{sw}$. The incompressible limit is easily recovered for $\rho\rightarrow\rho_0$, 
\begin{align}\label{exact_mhd}
	-\frac{4}{3}\varepsilon_I \ell &= \text{F}_I,
\end{align}
where F$_I$ is the projection of ${\bf F}_I=\rho_0\big\langle [(\delta\uh)^2+(\delta\textbf{B})^2]\delta\uh - 2(\delta\uh\cdot\delta\textbf{B})\delta\textbf{B} \big\rangle$ along the mean plasma flow velocity field. Equation \eqref{exact_mhd} corresponds to the exact relation for fully developed IMHD turbulence \citep[see,][]{P1998a,P1998b}. Here \textbf{B} is expressed in velocity units and $\varepsilon_I$ is the incompressible energy cascade rate. One can note that \textbf{F}$_I$ depends only on the increments of the magnetic (and velocity) field, although the total magnetic field has been considered in the derivation. Finally, assuming the Taylor hypothesis (i.e., $V\tau\equiv\ell$, where $V$ is the mean plasma flow speed), Eqs. \eqref{numlaw} and \eqref{exact_mhd} can be expressed as a function of time lags $\tau$.

\section{Solar wind observations and selection criteria}\label{sec:obs}

In order to analyze the solar wind turbulence at different heliocentric distances, our data intervals for PSP and THEMIS/MAVEN were divided into a series of samples of equal duration of 30 and 35 minutes, respectively. This particular time duration ensures having at least one correlation time of the turbulent fluctuations for each particular heliocentric distance \citep[see,][]{H2017a,Ma2018,P2020}. Moreover, we avoid intervals that contained significant disturbances or large-scale gradients (e.g., coronal mass ejection or interplanetary shocks). We further considered only intervals that did not show large fluctuations of the energy cascade rate over the MHD scales, typically we retained events with $\text{std}(\varepsilon_I)/\text{mean}(|\varepsilon_I|)<0.75$. 

\begin{table}[h!]
\begin{tabular}{||l c c c c||}
\hline
Mission & Magnetic-Plasma Instruments & Cadences [Hz] & Distances [AU] & $\#$ of samples \\ 
\hline\hline
PSP & FIELDS-SPC & 4.58 - 1.14 & $\sim$ 0.2 & 384 \\ 
\hline
THEMIS & FGM-ESA & 128 - 0.33 & $\sim$ 1.0 & 160 \\
\hline
MAVEN & MAG-SWIA & 32 - 0.25 & $\sim$ 1.6 & 104 \\
\hline
\end{tabular}
\caption{Short description of the data used.}
\label{table}
\end{table}

Table \ref{table} shows a brief description of the data used. Our analysis on the PSP observations involved two particular data sets: one covering the period between November 1 and November 10, 2018, which was dominated essentially by slow wind flows, and the other one between November 15 and November 21, 2018 dominated by high speed flows. In both sets, the spurious data (i.e., high artificial peaks) in the SPC moments were removed using a linear interpolation \citep[see,][]{B2020,P2020} and the data set was resampled to 0.873 s time resolution. For the data at 1 AU, we used the same data set analyzed in \citep{H2017a}, i.e., the period 2008-2011, where a large survey of the THEMIS data has been reported. This data set was resampled to 3 s time resolution and we only used the events that fulfilled our criteria, covering both the fast and slow solar wind. Finally, magnetic field observations from MAVEN spacecraft were analyzed to discriminate events in the pristine solar wind with and without wave activity \citep{R1990,M2004}. As discussed in the Introduction, recently, the incompressible energy cascade rate has been investigated at the Martian environment plasma \citep{A2020}. Here, we used the same data set for the solar wind observations without presence of proton cyclotron waves \citep[also see,][]{R2013,Ro2016,Ha2020,Ro2020}.

It is worth mentioning that the local solar wind temperature from the SWIA instrument (onboard MAVEN) and the ESA instrument (onboard THEMIS) may overestimate the temperature moments by a factor of $\sim$ 2 due to the trace presence of alpha particles \citep{Ha2017}. In fact, the local temperature enters in the compressible exact law only in the compressible term \eqref{numlaw} through the definition of the sound speed in the internal energy, i.e., $e = c_s^2 \log(\rho/\rho0)$, where $c_s$ is the isothermal sound speed. To investigate the impact of a possible overestimation of the solar wind temperature in the MAVEN observations, we have computed the compressible cascade component taking into account half of the measured temperature. While in some cases the compressible component decreases (by a factor $\sim$ 2), the statistical results, trends and conclusions in this paper are not affected by this potential temperature overestimation. 

\begin{figure*}
\begin{center}
\includegraphics[width=0.5\textwidth]{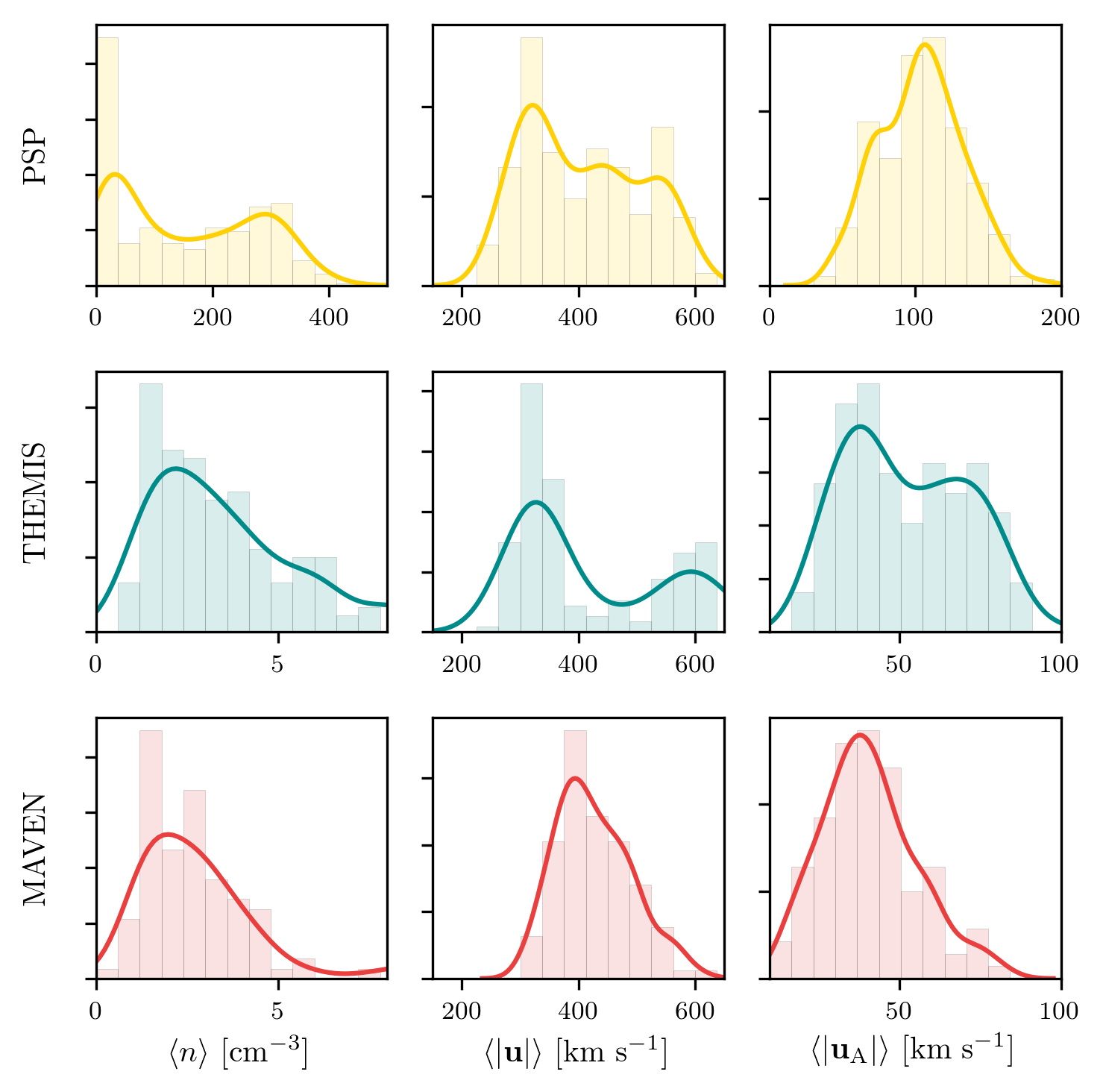}
\end{center}
\caption{For each spacecraft mission, the occurrence rate for the proton density, the proton and Alfv\'en velocity absolute values, respectively.}
\label{histograms}
\end{figure*}

Figure \ref{histograms} shows the occurrence rate for all the analyzed events in the three data sets for the number density, velocity, and Alfv\'en velocity field absolute values, respectively. As we expect, while THEMIS and MAVEN data sets have similar range of mean values for the number density, proton and Alfv\'en velocity absolute values, the PDFs for the PSP observations show clear increases in the number density (two order of magnitude) and the Alfv\'en velocity.

\section{Results}\label{sec:res}

\subsection{The compressible cascade rate and the MHD scales}

\begin{figure*}
\begin{center}
\includegraphics[width=0.5\textwidth]{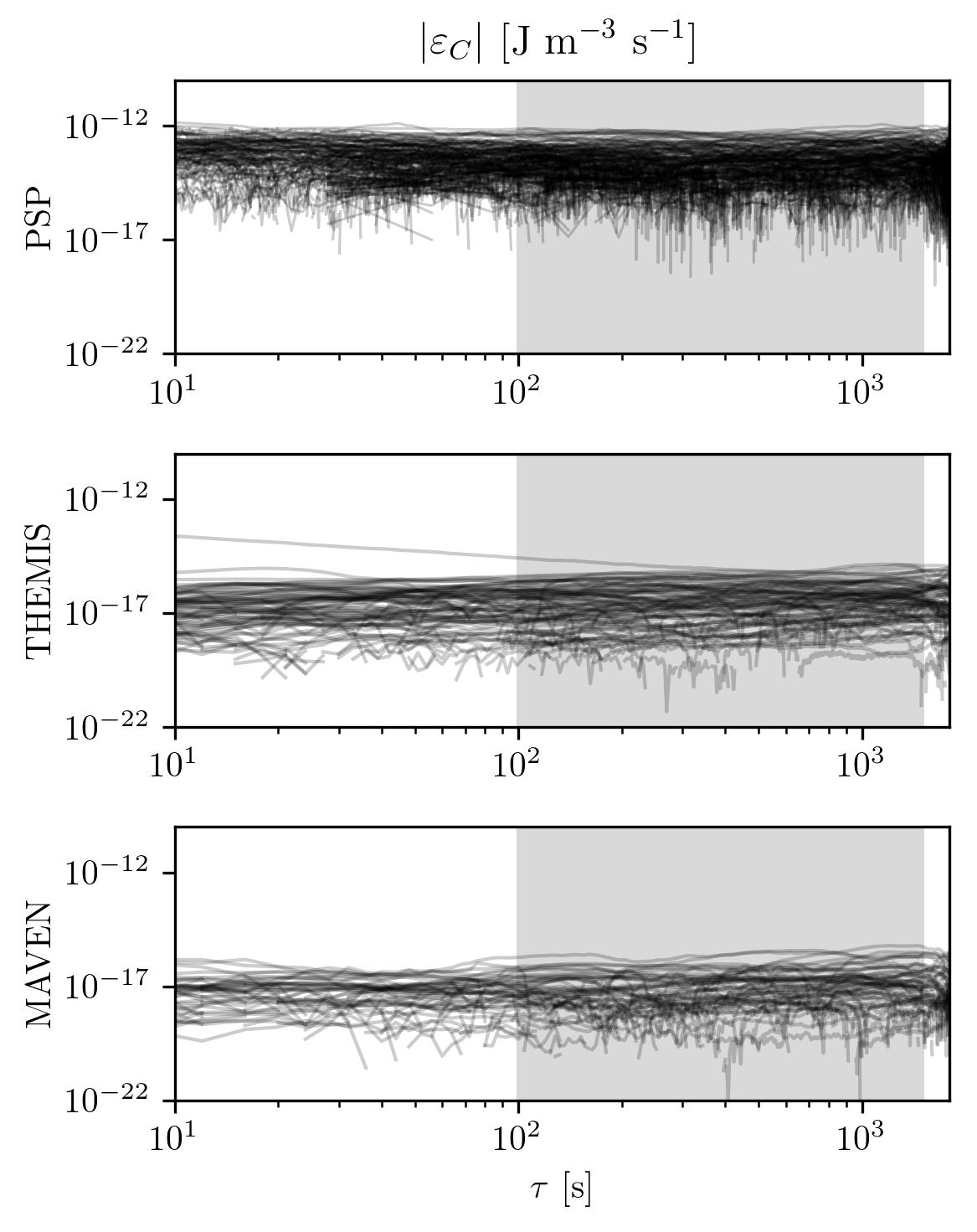}
\end{center}
\caption{Compressible energy cascade rates (absolute values) as a function of time lags for the three data sets. The MHD scales are showed in light gray background.}
\label{cascades}
\end{figure*}

In order to compute the compressible cascade \eqref{numlaw}, we constructed temporal correlation functions of the different turbulent fields at different time lags in the interval [4,2100] s. Figure \ref{cascades} shows the compressible energy cascade rate absolute values $|\varepsilon_C|$ as a function of the time lag $\tau$ for each mission. As a reference, in gray color mark what would correspond to the MHD inertial range. As we expect, our findings show a clear amplification of the compressible cascade as we approach to the Sun, which can be explained by the increase of  the magnetic and velocity field fluctuations closer to the sun (see Fig.~\ref{main}). In order to quantify statistically the increment in the cascade rates, we consider the mean of the absolute value in the largest MHD scales ($\tau\in[100,1800]$ s) as a representative value of each event.

\subsection{Compressible and incompressible energy cascade rates}

\begin{figure*}
\begin{center}
\includegraphics[width=0.65\textwidth]{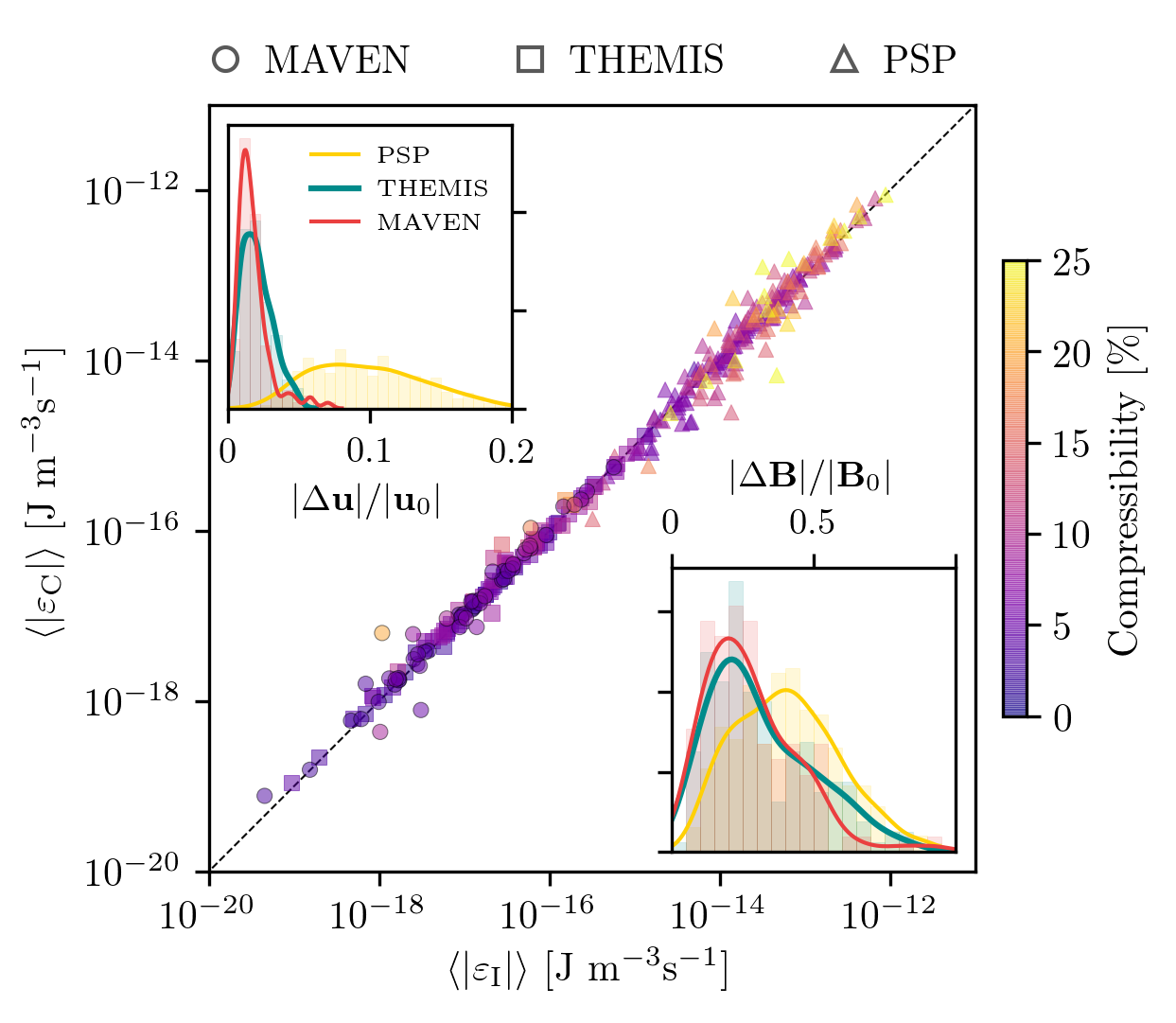}
\end{center}
\caption{Compressible energy cascade rate (absolute values) in the MHD scales as a function of the incompressible ones for each data set. Colorbar corresponds to the level of compressibility per event. Inset: histograms for the velocity and magnetic (absolute values) fields fluctuations.}
\label{main}
\end{figure*}

\begin{figure*}
\begin{center}
\includegraphics[width=0.3\textwidth]{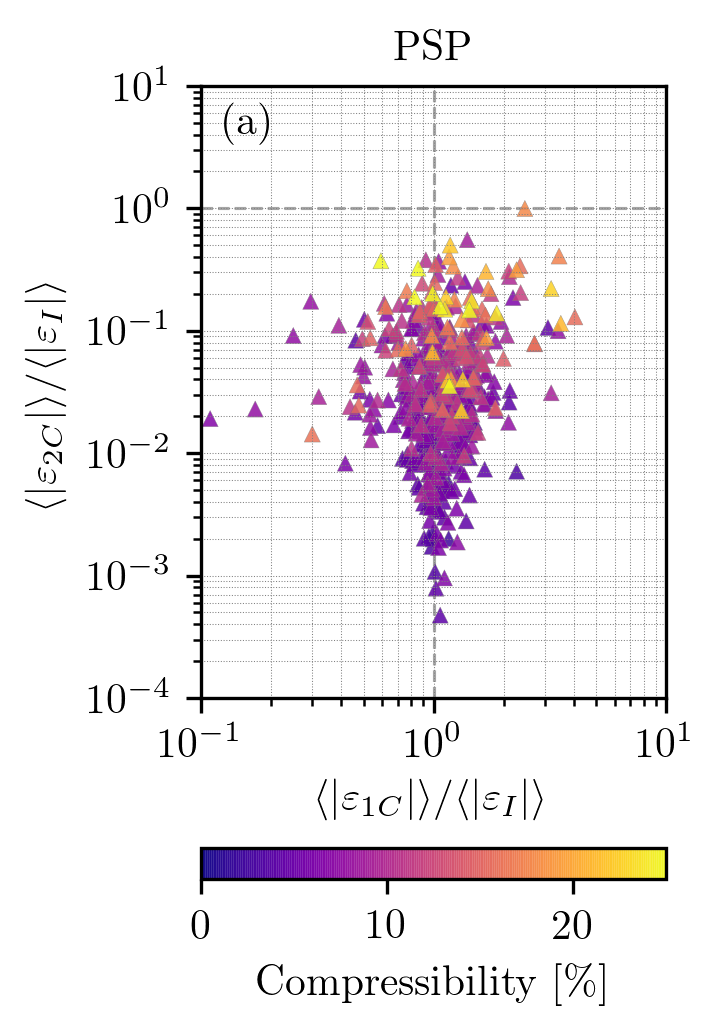}
\includegraphics[width=0.3\textwidth]{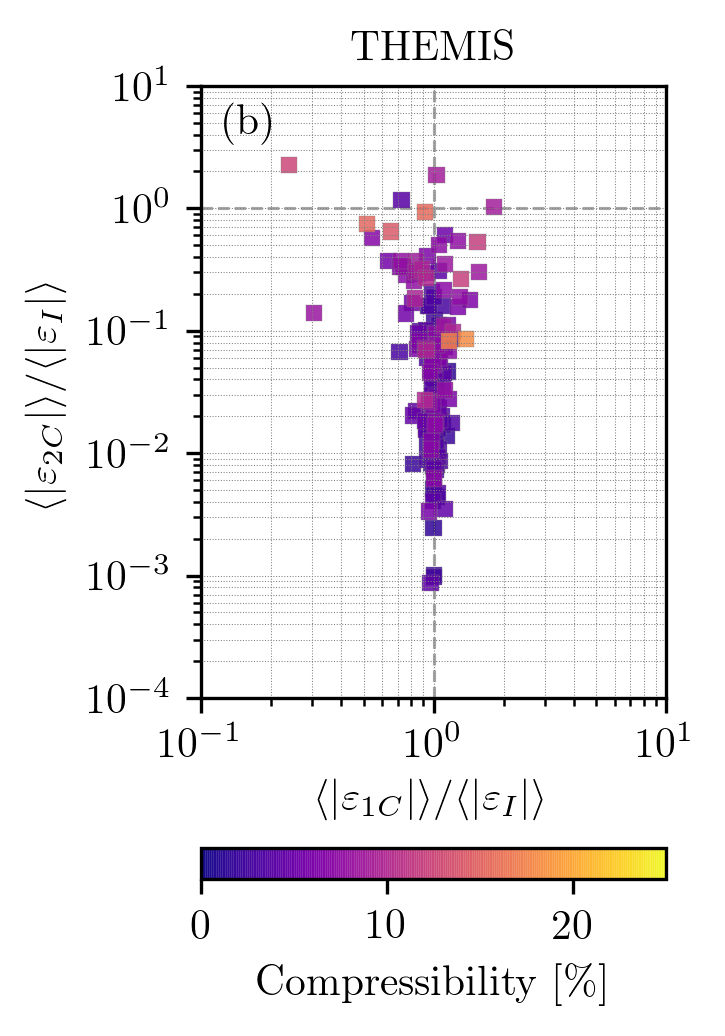}
\includegraphics[width=0.3\textwidth]{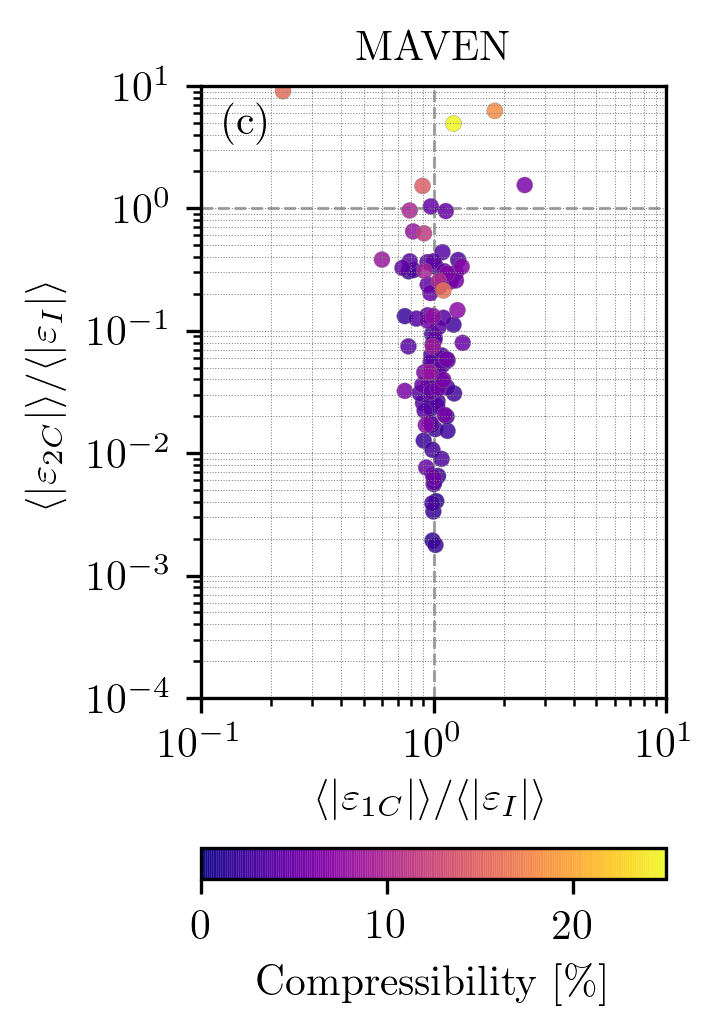}
\end{center}
\caption{The compressible cascade rate component $\langle|\varepsilon_\text{2C}|\rangle$ as a function of the Yaglom-like component $\langle|\varepsilon_\text{1C}|\rangle$, both normalized to the total incompressible component $\langle|\varepsilon_I|\rangle$. For the three missions, the color bar indicates the mean level of density fluctuations}
\label{ratios1}
\end{figure*}

\begin{figure*}
\begin{center}
\includegraphics[width=0.3\textwidth]{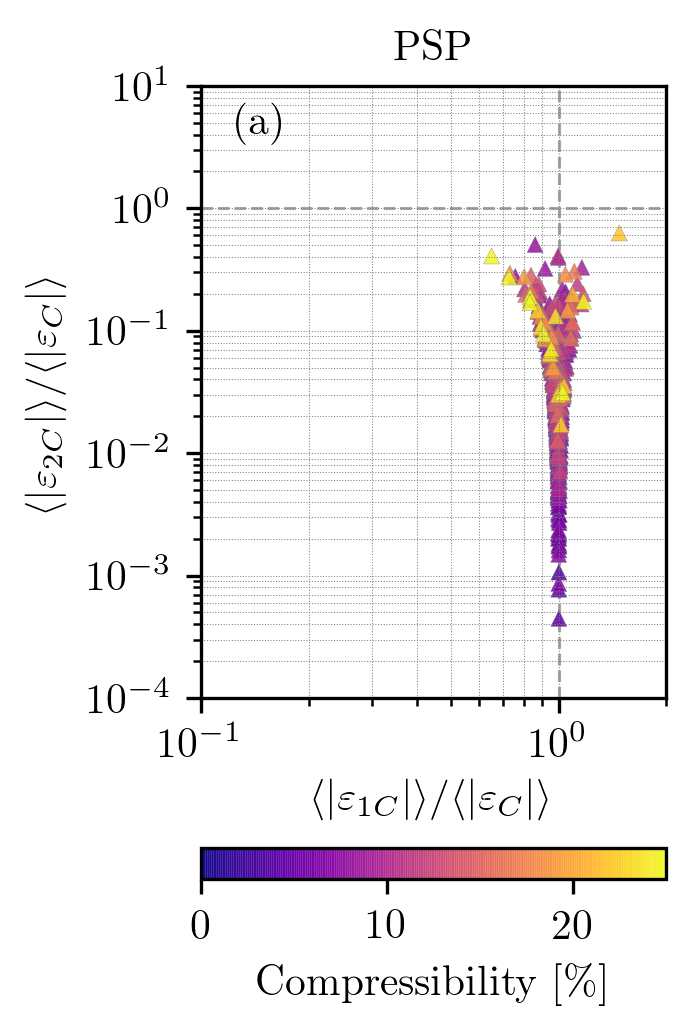}
\includegraphics[width=0.3\textwidth]{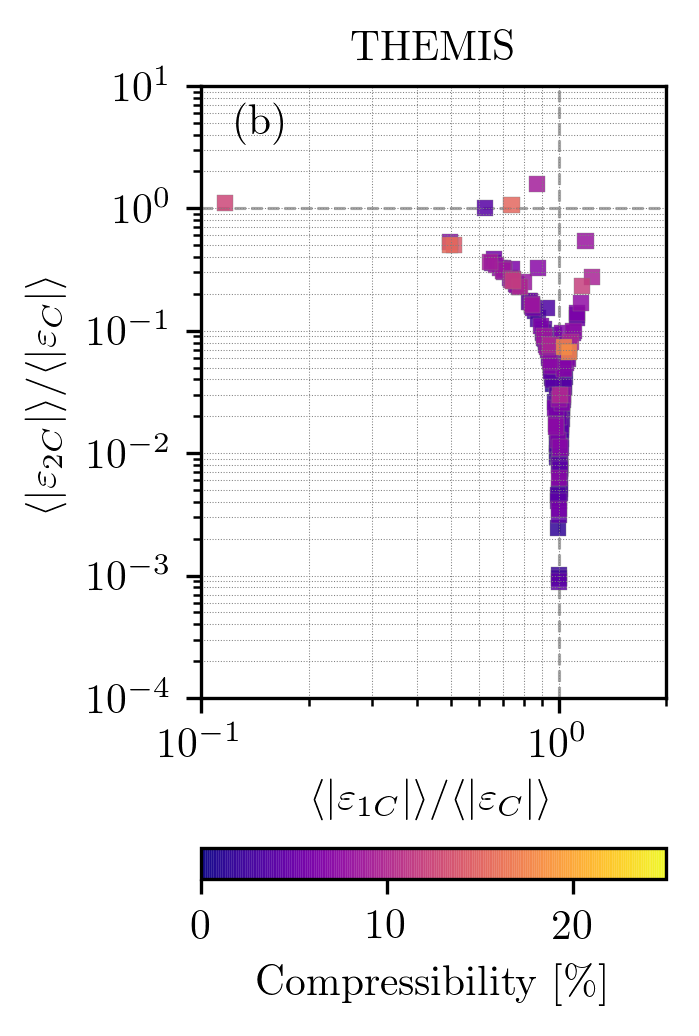}
\includegraphics[width=0.3\textwidth]{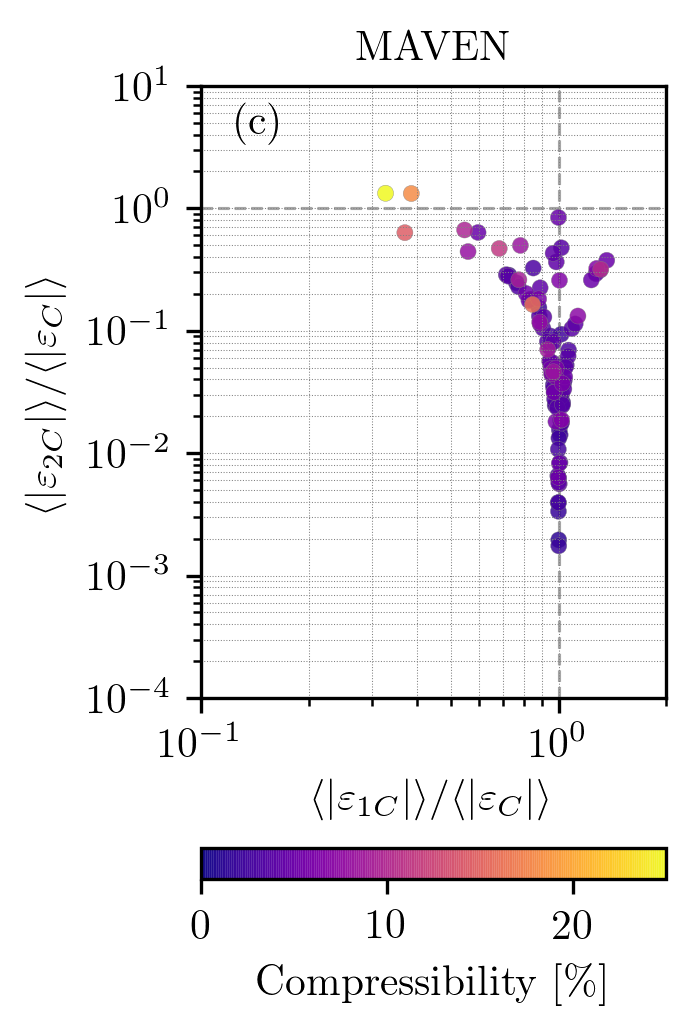}
\end{center}
\caption{The compressible cascade rate component $\langle|\varepsilon_\text{2C}|\rangle$ as a function of the Yaglom-like component $\langle|\varepsilon_\text{1C}|\rangle$, both normalized to the total compressible component $\langle|\varepsilon_\text{C}|\rangle$. For the three missions, the color bar indicates the mean level of density fluctuations}
\label{ratios2}
\end{figure*}

Figure \ref{main} shows $\langle|\varepsilon_\text{C}|\rangle$ as a function of $\langle|\varepsilon_\text{I}|\rangle$ for PSP (triangles), THEMIS (squares) and MAVEN (circles) observations. The colorbar represents the compressibility (in percent) of each event, defined as $\langle|n_i-n_0|\rangle/n_0$ (where $n_0$ is the mean number density). Insets correspond to the PDFs of the absolute values of the velocity and magnetic field fluctuations, respectively. In particular, the level of density, velocity and magnetic fluctuations increase as we approach to the Sun \citep[e.g.,][]{B2005,M2011}. On the other hand, considering that the plasma compressibility increases up to 25 $\%$ in the PSP perihelion, we observe moderate increases of the total compressible cascades with respect to the incompressible cascades. This is in agreement with previous work in the solar wind and the Earth's magnetosheath \citep{H2017a,H2019,A2019b}. 
%In particular, as we show below, the different terms ${\bf F}_1$ and ${\bf F}_2$ involve in the exact law \eqref{numlaw} may have different impact in the total compressible cascades according to the level of compressibility in the plasma. 

Figure \ref{ratios1} and \ref{ratios2} show the compressible component $\langle|\varepsilon_\text{2C}|\rangle$ (relate to Eq.~\eqref{f2}) as a function of the Yaglom generalization component $\langle|\varepsilon_\text{1C}|\rangle$ (relate to Eq.~\eqref{f1}), normalized to the total incompressible and compressible cascade rates, respectively. When the density fluctuations are small ($\le 5\%$), the compressible Yaglom-like component $\varepsilon_\text{1C}$ coincides with the incompressible component $\varepsilon_I$, while the compressible term $\varepsilon_\text{2C}$ is negligible. However, when the density fluctuations increase in the plasma (up to 25$\%$), the dominant component is given by a competition between $\varepsilon_\text{1C}$ and $\varepsilon_\text{2C}$. We recall that, independently of the level of density fluctuations, we assumed that the source, hybrid and $\beta$-dependent terms are negligible in the MHD inertial range based on the simulation results of \citet{A2018b}, and thus are not estimated here.

\subsection{Relation between cascade rate and local temperature}

\begin{figure*}[h!]
\begin{center}
\includegraphics[width=0.5\textwidth]{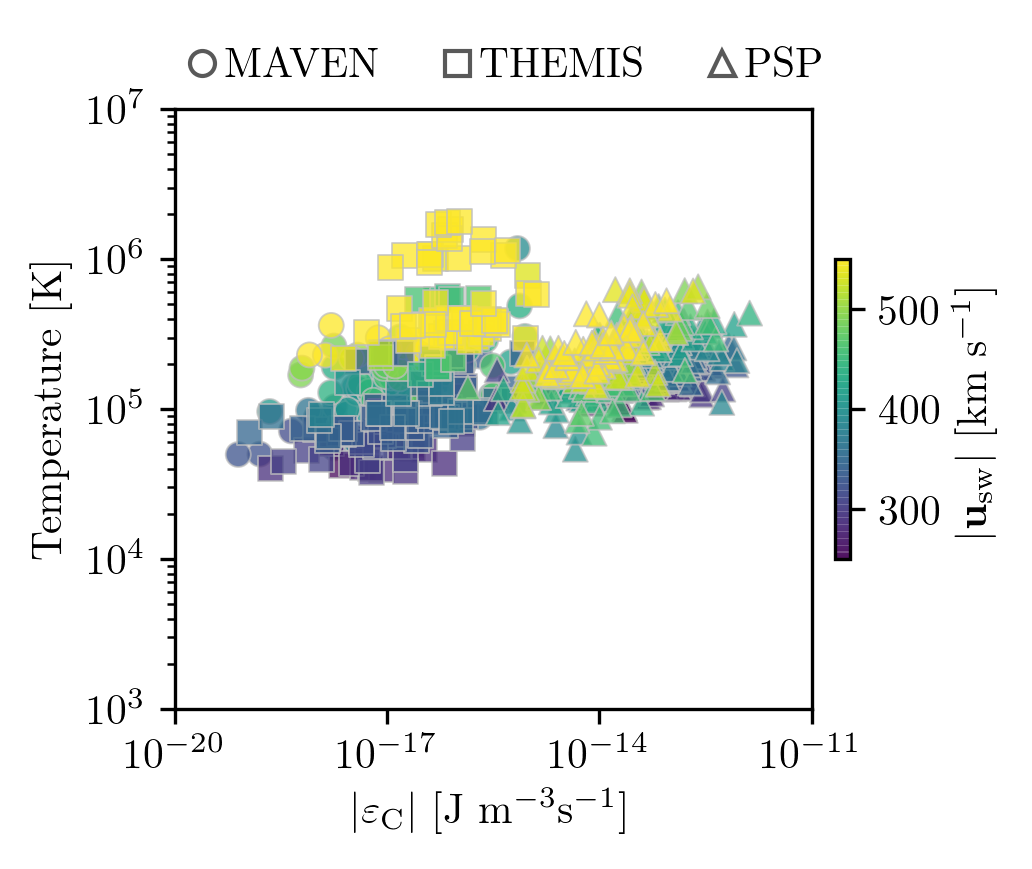}
\end{center}
\caption{The solar wind temperature as a function of the absolute value of the total compressible energy cascade rate $|\varepsilon_\text{C}|$. The colorbar corresponds to the mean solar wind speed.}
\label{rate_temp}
\end{figure*}

In order to connect the turbulent cascade rate and the local ion temperature, we study the correlation of the compressible cascade rate with the ion temperatures for the slow and fast solar wind plasma. Figure \ref{rate_temp} shows the solar wind ion temperatures as a function of the absolute value of the total compressible energy cascade rates. The colorbar corresponds to the mean solar wind velocity. The observational results show a clear trend between the increase in the solar wind temperature and the compressible cascade rates. As the cascade rate increases, we observe a slight increase in the local ion temperature. Note that the temperature increase also correlated with the increase in the solar wind speed.

%\newpage
\section{Discussions and Conclusions}

To the best of our knowledge, we reported the first estimation of the compressible energy transfer rate in the solar wind at $\sim$ 0.2 - 0.4 AU and 1.5 - 1.7 AU, using PSP and MAVEN observations, respectively. As we get close to the Sun, we observed a slight increase in the density, velocity and magnetic fluctuations \citep{B2005}. For the most compressible events (when compressibility is $\sim 25\%$), in the first PSP perihelion (on November 6, 2018), we observe that the compressible energy cascade rate is at least 5 order of magnitude larger than the cascade at $\sim$ 1.6 AU. Also, we observe a growth in the compressible cascade rate with respect to the incompressible rate. The increase in the compressible (and incompressible) cascade rates when we compare the results between 0.2 AU and 1.7 AU is mainly due to the increase in the magnetic and the number density fluctuation levels (see Figures \ref{histograms} and \ref{main}). Previous studies showed that the density fluctuations would increase considerably the cascade rate when compressibility is larger than $30 \%$ or when it enters to the sub-ion scales  \citep[see,][]{H2017a,A2018b,A2019b}. Our results are compatible with these findings at the MHD scales. Also, \citet{Ad2020} have computed the frequency distributions of the solar wind compressibility between 0.17 AU and 0.61 AU using PSP observations. Their results show that density fluctuations (normalized to the mean density) is concentrated mainly around 0.15, which is compatible with the density fluctuation levels found in this study. In contrast, recent computations of the magnetic compressibility coefficient $C_B \equiv (\delta|{\bf B}|/|\delta{\bf B}|)^2$ at 0.17 AU have shown a clear decreases toward smaller heliocentric distances (at all frequencies) \citep[see,][]{Ch2020}. In particular, the authors have observed that magnetic compressibility levels at the PSP perihelion are an order of magnitude smaller than at 1 AU. However, this decrease in the magnetic compressibility shows no overall impact in the density fluctuation level and consequently in the compressible cascade rate estimations.

Our findings here for different heliocentric distances confirm that density fluctuations in the solar wind amplifies the cascade rate with respect to the incompressible model \citep{SV2007,A2020,B2020}. When density fluctuations are relatively larger than $\sim15\%$, our results show that the leading role in the amplification of the cascade rate is due to a competition between the new compressible flux component $\langle|\varepsilon_\text{2C}|\rangle$ and the compressible Yaglom-like generalization $\langle|\varepsilon_\text{1C}|\rangle$ in the exact relation. As the plasma compressibility locally increases through the heliosphere we observe two clear features: on one hand, the Yaglom-like component $|\varepsilon_\text{1C}|$ becomes more spread (larger and/or smaller values) around the incompressible value. While this feature could be due to the relatively poor statistics in the THEMIS and MAVEN analyzed data sets, the spread could be due to the large plasma density fluctuation levels present in the PSP data set. On the other hand, as the plasma compressibility increases we observe that the compressible component $\varepsilon_\text{2C}$ becomes of the order of the Yaglom-like component in the inertial range. \citet{A2018b} have computed all the terms in Eq.~\eqref{exactlaw} using DNS for sub-sonic isothermal compressible MHD turbulence. The authors found that even for large guide field values, the energy cascade rate $\varepsilon_\text{C}$ is mainly due to the flux terms. Also, they observed that $\varepsilon_\text{2C}$ increases at least one order of magnitude when plasma compressibility is increased by a factor two in the system. Furthermore, they found that when compressibility increases in the plasma, the compressible component becomes dominant in the transfer of the energy. These numerical finding are in agreement with our observational results in Figures \ref{ratios1} and \ref{ratios2}. We speculate that this trend should be observed in more compressible plasma environments, like the Earth's magnetosheath \citep{Hu2017,H2017a,A2019b,L2020}.

Our observational results show a correlation between the solar wind temperature and the compressible energy cascade rate. The larger is the cascade rate, the higher is the temperature. Also, a clear increase in the temperature is connected with fast solar wind speeds. The relation between the cascade rate and the local temperature in the solar wind has been studied previously in the literature \citep[see,][]{V2007,Mc2008,M2008,B2016c}. As we discussed above, the turbulent cascade rate has been proposed as a solution to the solar wind heating problem \citep[e.g.,][]{R1995,M2011}. Our observational results here support the idea that the compressible turbulent cascade may heat the plasma at different heliocentric distances. Finally, we emphasize that all results reported here were obtained inside the solar wind ecliptic plane. A natural extension of this study would be the use of the ESA recently launched Solar Orbiter mission \citep{M2020} data. Its observations should give us access to the polar regions of the solar wind within heliocentric distance never explored before. 

%Our results complement these compressible and incompressible results, showing how different cascade components are affected due to density fluctuations. Finally, the dependence of the compressible cascade respect to the presence of cyclotron waves or a potential seasonal variability dependence around Mars environment is beyond the scope of the present work.

\newpage 
\section*{Acknowledgements}
N.A. acknowledges financial support from the following grants: PICT 2018 1095 and UBACyT 20020190200035BA. N.A., L.H.Z., and F.S. acknowledge financial support from CNRS/CONICET Laboratoire International Associé (LIA) MAGNETO. N.R. is an Assistant Research Scientist at NASA Goddard Space Flight Center, hired by MAVEN Project Scientist Support, and administered by CRESST II and UMBC. We thank the entire PSP, THEMIS and MAVEN team and instrument leads for data access and support. PSP data are publicly available (\url{https://research.ssl.berkeley.edu/data/psp/}). The THEMIS/ARTEMIS data come from the AMDA
database (\url{http://amda.cdpp.eu/}). MAVEN data are publicly available through the Planetary Data System (\url{https://pds-ppi.igpp.ucla.edu/index.jsp}). 

\bibliography{cites}{}
\bibliographystyle{aasjournal}

\end{document}